\documentclass[
    ,final            
  ]
  {aipproc}

\def\msun{$M_{\odot}$}

\def\risco{$R_{\rm ISCO}$}

\layoutstyle{6x9}

\begin{document}

\title{Measuring the Spins of Stellar Black Holes: \\ A Progress Report}


\author{J.\ E.\ McClintock, R.\ Narayan, L.\ Gou, J.\ Liu, R.\ F.\ Penna
  and J.\ F.\ Steiner}{ address={Harvard-Smithsonian Center for
  Astrophysics, 60 Garden St., Cambridge, MA 02138 USA} }

\begin{abstract}
We use the Novikov-Thorne thin disk model to fit the thermal continuum
X-ray spectra of black hole X-ray binaries, and thereby extract the
dimensionless spin parameter $a_* = a/M$ of the black hole as a
parameter of the fit.  We summarize the results obtained to date for six
systems and describe work in progress on additional systems.  We also
describe recent methodological advances, our current efforts to make our
analysis software fully available to others, and our theoretical efforts
to validate the Novikov-Thorne model.
\end{abstract}

\maketitle


\section{Introduction}

We now know of about 40 stellar-mass black holes (BHs) in X-ray binaries
in the Milky Way and neighboring galaxies with masses ranging from $\sim
5-20$\msun~[1]. Astrophysical BHs are completely described by the two
numbers that specify their mass and spin.  BH spin is commonly expressed
in terms of the dimensionless quantity $a_* \equiv cJ/GM^2$ with $|a_*|
\le 1$, where $M$ and $J$ are respectively the BH mass and angular
momentum.  Currently, there are two techniques that are delivering
measurements of spin, namely fitting the thermal X-ray continuum [2--6]
and modeling the profile of the Fe K line.  Our group is engaged in
using the continuum-fitting (CF) method, which is the focus of this
paper.  (For a discussion of the Fe K method see [7]).

Knowledge of BH spin is crucial for answering many key questions.  For
example: Are relativistic jets powered by spin?  What role does spin
play in producing a gamma-ray burst?  What constraints can be placed on
models of supernovae, BH formation, and BH binary evolution?  What
distribution of BH spins should LIGO waveform modelers be considering?
For supermassive BHs, is the distribution of spins of the merging
partners consistent with hierarchical models for their growth?

In the following two sections, we describe the CF method of determining
spin and present the results we have obtained to date, while describing
our current work on four additional sources.  The next section describes
recent methodological advances, including our efforts to make all of our
fitting software publicly available.  The penultimate section describes
our work aimed at validating the theoretical underpinning of our work --
the Novikov-Thorne disk model -- via GRMHD simulations.  We conclude
with a number of questions that motivate us.

\section{The Method: Fitting the X-ray Continuum Spectrum}

Here we present the bare elements of the method; for a fuller
explanation, see Sec.\ 2 in [9].  The foundation of the CF method is the
definite prediction of relativity theory that there exists an innermost
stable circular orbit (ISCO) for a test particle orbiting a BH.  We
identify the radius of the ISCO, $R_{\rm ISCO}$, with the inner edge of
the accretion disk.  Thus, the gas spirals in (through the action of
viscosity) via a series of nearly circular orbits until it reaches the
ISCO, at which point it plunges into the BH.  In our method, we estimate
the radius of the inner edge of the disk by fitting the X-ray continuum
spectrum and identify this radius with $R_{\rm ISCO}$.  Since the
dimensionless ratio $\xi \equiv R_{\rm ISCO}/(GM/c^2)$ is solely a
monotonic function of the BH spin parameter $a_*$, knowing its value
allows one immediately to infer the BH spin parameter $a_*$.  The
variations in \risco~are large, ranging from $6GM/c^2$ to $GM/c^2$ as 
$a_*$ increases from 0 to unity.

Thousands of observations of BH binaries in the thermal state, which
have been made during the past 25 years, suggest that fitting the X-ray
continuum is a promising approach to measuring BH spin.  Specifically,
these observations have provided abundant evidence for a constant inner
disk radius in the thermal state.  For discussions of this empirical
foundation of the method see [6,10].

The idealized thin disk model of Novikov \& Thorne [8] describes an
axisymmetric radiatively-efficient accretion flow in which, for a given
BH mass $M$, mass accretion rate $\dot M$ and BH spin parameter $a_*$,
we can calculate precisely the total luminosity of the disk, $L_{\rm
disk} = \eta\dot Mc^2$, where the radiative efficiency factor $\eta$ is
a function only of $a_*$.  Moreover, the accreting gas is optically
thick, and the emission is thermal and blackbody-like, making it
straightforward to compute the spectrum of the emission and other
properties of the disk, such as its luminosity profile $L(R)$.  Most
importantly, as discussed above, the inner edge of the disk is located
at the ISCO of the BH space-time.  By analyzing the spectrum of the disk
radiation and combining it with knowledge of the distance $D$, the
inclination angle $i$ and the mass $M$ of the BH, we can obtain $a_*$.
This is the principle behind our method of estimating BH spin, which was
first described by Zhang et al.\ [11].

For the CF method to succeed it is essential to have accurate
measurements of the BH mass $M$, inclination of the accretion disk $i$,
and distance $D$ as inputs to the continuum-fitting process [2,3].  This
dynamical work is not discussed here, although about half of our total
observational effort is directed toward securing these dynamical data
(e.g., see [12,13]).

\section{Results to Date and Work in Progress}

In the top portion of Table 1, we summarize the half-dozen measurements
of spin obtained using the CF method during the past four years.  The
values of spin range widely: LMC X-3 is a near Schwarzschild hole and
GRS 1915+105 is a near-extreme Kerr hole, while four sources have
intermediate values of spin in the range $a_* \sim 0.7-0.9$.  The values
of BH mass used in deriving the spin estimates are given in the table.
For the details on this published work we refer the reader to the papers
cited, and we now turn to comment on the four sources listed at the
bottom of the table.

{\it XTE J1550-564:} Extensive photometric and spectroscopic data have
been collected, and a paper on a new dynamical model of the system is in
preparation, which will supersede our earlier model [15].  Meanwhile, a
preliminary spin analysis of 136 {\it RXTE} spectra has already been
published [16], and we are eagerly working toward a definitive
measurement of the spin of this pc-scale ballistic jet source.

\begin{table}[!t]
\begin{tabular}{rrrp{.35\textwidth}} 
\hline
  \tablehead{1}{l}{b}{BH Binary System} &
  \tablehead{1}{c}{b}{$M/M_{\odot}$} & 
  \tablehead{1}{c}{b}{$a_*$} &
  \tablehead{1}{l}{b}{References} \\
\hline
4U 1543--47   & $9.4\pm1.0$     & 0.75--0.85              & [2] \\
GRO J1655--40 & $6.30\pm0.27$   & 0.65--0.75              & [2] \\
GRS 1915+105  & $14\pm4.4$      & 0.98 - 1                & [3] \\
LMC X-3       & $7 (5-11)$      & $<0.26$                 & [4] \\
M33 X-7       & $15.65\pm1.45$  & $0.77\pm0.05$           & [5,12] \\
LMC X-1       & $10.91\pm1.41$  & $0.92_{-0.07}^{+0.05}$  & [6,13] \\
XTE J1550-564 & TBD             & TBD                     & Steiner et al.; Orosz et al. \\
A0620--00     & TBD             & TBD                     & Gou et al.; [14] \\
Cygnus X-1    & TBD             & TBD                     & Gou et al.; Orosz et al.; Reid et al. \\
LMC X-3       & TBD             & TBD                     & Steiner et al.; Orosz et al. \\
\hline
\end{tabular}
\caption{Spin Measurements of Stellar Black Holes: Published and in the Works}
\label{tab:b}
\end{table}

{\it A0620-00:} This system, which brightened to an unprecedented 50
Crab in 1975, is the prototype of the nine short-period BH binaries
($P_{\rm orb}<12$ hr) [1].  Using a new determination of the BH mass,
inclination and distance, which is based on an exhaustive study of the
available photometric data [14], we are in the process of determining the
spin using HEASARC archival data obtained in 1975 by the OSO-8
satellite.

{\it Cygnus X-1:} Within several months, we expect to have an accurate
VLBA parallax measurement of the distance to this source (Reid et al.),
which will be followed by dynamical modeling and a determination of the
BH mass and inclination.  Meanwhile, we are working to determine the
spin using {\it ASCA} GIS data and {\it RXTE} PCA data, which were
obtained simultaneously [17].

{\it LMC X-3:} This source appears twice in Table 1.  The spin
constraint given in line 4 is not firm because it is based on an early
and uncertain estimate of the BH mass [4].  We have recently obtained
extensive and very high quality spectroscopic and photometric data and
are in the process of revisiting the dynamical model.  An extensive spin
analysis of the X-ray data has already been completed [18].

\section{Recent Advances in Methodology}

As we now describe, using new techniques we are able to both
successfully apply the CF method to a larger body of data and to derive
uncertainties in the spin parameter that include all of the
observational uncertainties.  In this section we also describe our
tentative plans for improving our relativistic code and making it fully
accessible via XSPEC.

\vspace{2.mm}

{\bf Beyond the Thermal Dominant State:}  All of our prior work on
measuring spins has relied on the use of weakly Comptonized spectra
obtained in the thermal dominant (TD) state.  Now, however, using our
recently-developed empirical model of Comptonization {\sc simpl} [19],
which is available in XSPEC, we are able to obtain values of spin that
are consistent with those obtained in the TD state.  We have
demonstrated this capability by analyzing many {\it RXTE} spectra of two
BH transients, H1743-322 and XTE J1550-564, and showing that the radius
of the inner edge of their accretion disks remains constant to within a
few percent as the strength of the Comptonized component increases by an
order of magnitude, i.e., as the fraction of the thermal seed photons
that are scattered approaches 25\% [16].  This development allows us to
apply the CF method to a much wider body of data than previously thought
possible, and potentially to sources that have never been observed to
enter the TD state.

\vspace{2.mm}

{\bf Monte-Carlo Error Analysis:} In early work [2,3], we made only
quite approximate estimates of the error in the spin parameter.  In
contrast, the error analysis in our most recent papers on M33 X-7 and
LMC X-1 is much more sophisticated [5,6].  Therein, we determine the
error in $a_*$ due to the combined uncertainties in $M$, $i$, and $D$
via Monte Carlo simulations assuming that the uncertainties in these
parameters are normally and independently distributed.  Most recently,
for LMC X-1 we also performed a combined error analysis that considers
both of our fiducial values of the viscosity parameter (see Fig.\ 8 in
[6]); thus the error in this case includes the uncertainty in this key
model parameter as well as all sources of observational error.
Meanwhile, the largest error in our results arise from uncertainties in
the validity of the disk model we employ (see below).  Another source of
uncertainty is our assumption that the spin of the BH is aligned with
the orbit vector to within a few degrees.  This question of relative
alignment will be addressed directly through observations using the GEMS
mission X-ray polarimeter now scheduled for flight in 2014 [20,21].

\vspace{2.mm}

{\bf Improvements in Relativistic Disk Codes and Public Access:} Our
workhorse accretion disk model is available in XSPEC under the name {\sc
kerrbb} [22].  It includes all relativistic effects and additional
features; most importantly it includes self-irradiation of the disk
(``returning radiation'').  A limitation of {\sc kerrbb} is that one of
its three key fit parameters, namely, the spectral hardening factor $f$,
is treated as a constant.  Because of this limitation, our work is also
based on a second, complementary relativistic disk model called {\sc
bhspec} [23,24], which is also implemented in XSPEC.  The model {\sc
bhspec} does not include the effects of returning radiation, but it does
provide state-of-the-art capability for computing the spectral hardening
factor $f$.  In all of our work since Shafee et al. [2], we have used a
hybrid code that combines the functionalities of {\sc bhspec} and {\sc
kerrbb} into a single code we call {\sc kerrbb2} (see Sec.\ 4.2 of [3]
for details).

It is quite awkward to analyze data using the hybrid code {\sc kerrbb2}
because for each fit one must first use {\sc bhspec} to generate a large
multi-dimensional table of the spectral hardening factor $f$, which
depends on the relevant detector response function, and then read values
of $f$ from this table while performing the fit using {\sc kerrbb}.  We
are finding that it is quite difficult to implement {\sc kerrbb2} for
public use within XSPEC.  Consequently, we are now developing a new and
simpler hybrid code {\sc bhspec2} that uses {\sc bhspec} as the primary
engine and includes the effects of returning radiation computed using
{\sc kerrbb}.  The significant advantage of this approach is that it is
independent of detector response and is therefore simpler.  Our goal is
to make {\sc bhspec2} publicly available in XSPEC during the first half
of 2010.  Our further, three-year goal is to create a new version of
{\sc bhspec} that organically includes returning radiation.

\section{Testing the Novikov-Thorne Model}

Any measurement of BH spin is only as good as the theoretical model
behind it.  The CF method assumes that the radial luminosity profile of
the disk $L(r)$ is given by the analytical form derived by Novikov \&
Thorne [8].  However, the validity of the NT model, and in particular
the zero-torque boundary condition at the ISCO, which it assumes, has
been questioned [25,26].  Because any serious error in the NT model will
lead to large systematic errors in the derived BH spin values, we have
mounted a major effort to scrutinize the NT model.

We are carrying out GRMHD simulations of thin accretion disks in the
Kerr metric and comparing the simulation results with the predictions
of the NT model.  Our first results for a nonspinning BH are reported
in Shafee et al. [2] where we show that, for a disk with a
dimensionless thickness parameter $H/R \sim 0.05$, there is little
evidence for significant magnetic coupling across the ISCO.  In
particular, the angular momentum profile of the simulated flow agrees
very closely ($<2$\% difference) with the NT prediction.  In ongoing
work, we are simulating disks of a variety of thicknesses, $H/R \sim
0.05$, 0.1, 0.2, 0.3, around BHs of various spins, $a_*=0$, 0.7, 0.9,
0.98.  Once again we find that, for all four values of $a_*$, the
thinnest disk models closely resemble the NT model in terms of their
angular momentum profiles, whereas thicker disks show progressively
larger deviations.  This is an extremely encouraging result for our BH
spin program.  All our spin estimates to date have been obtained using
X-ray spectral data on relatively low-luminosity systems with $L_{\rm
disk}/L_{\rm Edd} \leq 0.3$.  At these luminosities, the disks are
estimated to have $H/R \leq 0.1$ [3], suggesting that the spin
estimates we have reported so far are robust.

We are also extracting from the GRMHD simulations the luminosity
profiles $L(r)$ of the various disk models with a view to comparing
these with the predictions of the NT model.  The function $L(r)$,
which is the most crucial model input for the CF method, is
unfortunately less easy to estimate accurately from simulations; it
involves taking the difference of two large quantities, whereas the
angular momentum profile discussed earlier does not involve such a
difference.  Nevertheless, our current results are encouraging in the
sense that, for thinner disks, our numerically derived profiles of
$L(r)$ seem to agree well with the analytical $L(r)$ of the NT model.
All our spin estimates have been obtained using the latter profile, so
we continue to be confident in the spin estimates listed in Table 1.
Meanwhile, we are pushing ahead with additional state-of-the-art GRMHD
simulations of thin and thick disks and hope to have definitive
results shortly.

\section{Conclusion}

We conclude with a list of questions that motivate us as we work toward
our goal of measuring the spins of a dozen or more BHs.  What range and
distribution of spins will we find?  Will GRS 1915+105 stand alone, or
will we find other examples of extreme spin?  As we continue to refine
our models and our measurements of $M$, $i$ and $D$, will we
consistently find values of $a_* < 1$, or will we be challenged by
apparent and unphysical values of the spin parameter that exceed unity?
Will all our spin estimates be positive, or will we find some BHs with
$a_* < 0$, corresponding to a counter-rotating disk?  Will there be
large differences in spin between the class of young, persistent systems
with their massive secondaries (Cyg X-1, M33 X-7, LMC X-1 and LMC X-3)
and the ancient transient systems with their low-mass secondaries?  What
constraints will these spin results place on BH formation, evolutionary
models of BH binaries, models of relativistic jets and gamma-ray bursts,
etc.?  What will be the implications of these spin measurements for the
emerging field of gravitational-wave astronomy in the Advanced LIGO era?
How will this new knowledge help shape the observing programs of IXO,
LISA and other future space missions?




\end{document}